\documentclass[aps,prb,preprint,grouppedaddress,showpacs,showkeys]{revtex4}

\usepackage{graphicx}%
\usepackage{dcolumn}
\usepackage{amsmath}
\usepackage{multirow}
\usepackage{epsfig}
\usepackage{amssymb}
\usepackage{color}

\begin{document}

\title{Origin of doping-induced suppression and reemergence of magnetism in LaFeAsO$_{1-x}$H$_x$}

\author{Chang-Youn Moon}
\email{cymoon@kriss.re.kr}
\affiliation{Materials Genome Center, Korea Research Institute of Standards and Science, Yuseong, Daejeon 305-340, Republic of Korea}
\author{Hyowon Park}
\affiliation{Department of Physics, University of Illinois at Chicago, Chicago, Illinois 60607, USA}
\affiliation{Materials Science Division, Argonne National Laboratory, Argonne, Illinois 60439, USA}
\author{Kristjan Haule}
\affiliation{Department of Physics, Rutgers University, Piscataway, New Jersey 08854, USA}
\author{Ji Hoon Shim} \email{jhshim@postech.ac.kr}
\affiliation{Department of Chemistry, Pohang University of Science and Technology, Pohang 790-784,
Republic of Korea}

\date{\today}

\begin{abstract}
We investigate the evolution of magnetic properties as a function of hydrogen doping in iron
based superconductor LaFeAsO$_{1-x}$H$_x$ using the dynamical mean-field theory combined
with the density-functional theory. We find that two independent consequences of the doping,
the increase of the electron occupation and the structural modification, have the opposite
effects on the strength of electron correlation and magnetism, resulting in the minimum of
the calculated magnetic moment around the intermediate doping level as a function of $x$. 
Our result provides a natural explanation for the puzzling recent experimental discovery
of the two separated antiferromagnetic phases at low and high doping limits.
Furthermore, the increase of orbital occupation and correlation strength with the doping
results in reduced orbital polarization of $d_{xz/yz}$ orbitals and the enhanced
role of $d_{xy}$ orbital in the magnetism at high doping levels, and their possible implications 
to the superconductivity are discussed in line with the essential role of the magnetism.

\end{abstract}

%\keywords{silicon, nanowire, doping, magnetic interaction, density functional theory}

\maketitle

\section{Introduction}
%As-Fe-As angle decreases from 113 for x=0 to 108 for x=0.6.
Iron-based high temperature ($T$) superconductors
including the seminal material LaFeAsO$_{1-x}$F$_x$\cite{Kamihara}
share a common feature that the impurity doping results in the suppression of magnetic and/or
structural orders in undoped parent compounds and subsequent emergence of superconductivity.
%with a few exceptions such as LiFeAs \cite{Pitcher,Wang1} and FeSe \cite{Hsu} which exhibit
%the superconductivity without doping.
The underlying mechanism is still not well understood,
however, as the impurity doping has several independent effects on the system such as the change of
the electron occupancy, structural modification, and occurrence of the disorder, etc. One of
the popular explanations has been based on the itinerant picture of the antiferromagnetic (AFM)
ordering in undoped samples. As the doping changes the number of carriers and the position
of the Fermi level, the Fermi surface (FS) nesting condition for the spin density wave (SDW) formation
becomes poorer.\cite{Singh,Mazin,Cvetkovic,Dong} However, the non-negligible role of the electron
correlation in these materials has been pointed out,\cite{Haule1,Haule2,Yi} and the validity of
the FS nesting picture alone, which assumes the rigid band against the carrier doping, often turns out to be
doubtful in explaining the emergence and suppression of the AFM order. On the other hand, the structural
modification effect is known to alter significantly
the magnetic property with the strong magnetostructural coupling in this system.\cite{Singh,Yin1,Yildirim1}
Even the isovalent impurity doping or the hydrostatic pressure alone, which introduce no extra carrier,
can give rise to the similar phase diagram with the case of the carrier doping.\cite{Ren,Wang2,Park,Okada,Mukuda}

There is a general consensus about the importance of the magnetism in understanding the
superconductivity of iron-based materials,\cite{Hirschfeld,Lee1} as the magnetism is omnipresent in this class
of materials at least in the form of the short-range spin fluctuation.\cite{Li,Iikubo,Qiu} However, the
basic nature of the magnetism has been controversial among the SDW of itinerant electrons,\cite{Singh,Mazin,Cvetkovic,Dong,Dhaka}
the Heisenberg type interactions of localized spins,\cite{Si,Yang,Yildirim2,Zhao} as well as their intermediate
picture. \cite{Wu1,Wu2,Kou,Samolyuk,Zhao2,Moon1,Moon2,Ikeda,Yin2,PDai}
Nevertheless, the spin fluctuation has been widely accepted as the most probable candidate
for the pairing glue for the superconductivity. \cite{Mazin2,Zhang,Lee2,Ji,Dai,Yin2014,Allan} In the meanwhile,
there are also growing arguments and evidences for the presence of the orbital order and fluctuations
in these materials. \cite{Ku,Lv,Chen,WCLee1,WCLee2,Stanev,CKim,Ma,Liang,Kontani1,Saito,Mukherjee}
Recent experiments on FeSe suggest that the
orbital degree of freedom drives the electronic nematicity and spontaneous
symmetry breaking instead of the spin degree of freedom, \cite{Baek,Bohmer} drawing attention for
the alternative mechanism of the pairing mediated by the orbital fluctuation. \cite{Kontani2,Yanagi}
Because the spin and orbital degrees of freedom are coupled each other, \cite{Ku,Fernandes} however,
there can be an inherent ambiguity in determining which order forms first and drives the other.

Recently, a series of experiments has revived the attention to the seminal material of
the family by the hydrogen (H) doping, namely, LaFeAsO$_{1-x}$H$_x$.\cite{Iimura,Fujiwara,Hiraishi}
Overcoming the solubility limit of the conventional fluorine dopant, hydrogen can increase the
doping concentration up to $x=0.6$. Surprisingly, another superconducting and AFM phases
adjacent to each other are found at high doping levels, analogously to their conventional counterparts
at the low doping level (see Fig.~1(a)).
Posing fundamental questions on the nature of magnetism and
superconductivity, this finding is expected to give a clue for still unresolved issues
mentioned earlier. Although some theoretical attempts have been made to explain the appearance
of the second AFM phase mostly focusing on the FS nesting property in the itinerant electron
picture,\cite{Iimura,Fujiwara,Suzuki} first-principles approach simultaneously incorporating
the itinerant and localized aspects of the system is desirable when we consider the 'moderately correlated'
nature of these materials \cite{Haule1,Haule2,Yi,Lu,Qazilbash} and some unsatisfactory conclusions from the
itinerant picture such as the prediction of an incorrect magnetic ordering vector.\cite{Fujiwara}

In this paper, we investigate the magnetic and electronic properties of LaFeAsO$_{1-x}$H$_x$
as a function of $x$ using the combined method of density-functional
theory plus dynamical mean-field theory (DFT+DMFT), which captures the material-specific electronic
correlation.\cite{Kotliar,Haule3} Considering changes of both electron occupancy and lattice
structure caused by the hydrogen doping which turn out to have the opposite effects on the electron
correlation and magnetism, we find that both static magnetic moment and local
magnetic susceptibility initially decrease to the minimum at around $x=0.3$ and then 
increase again up to $x=0.6$, in agreement
with the experimental phase diagram of the two separate AFM phases centered at $x=0$ and 0.5.
%It is also found that the increasing Fe-As distance with the doping makes the system more
%localized at higher doping levels overcoming the counter effect of the electron addition
%away the half-filling, suggesting the second AFM phase at $x=0.5$ has more localized
%nature than the undoped case.
More electron occupation at $d_{xz/yz}$ orbitals with
the doping enhances the importance of the $d_{xy}$ orbital in the static magnetic moment
and also in spin dynamics, while reducing the orbital polarization. Our results emphasize
the importance of the electron correlation and structural modification in understanding the
doping induced evolution of the electronic structure, and also the magnetism as an
indispensable ingredient for the emergence of the superconductivity in these materials.

\section{Calculation Method}
We use the modern implementation of DFT+DMFT method within all electron embedded
DMFT approach,\cite{Haule3} where in addition to correlated Fe atoms the itinerant states of other
species are included in the Dyson self-consistent equation. The strong correlations on the Fe ion
are treated by DMFT, adding self-energy $\Sigma(\omega)$ on a quasi atomic orbital in real space,
to ensure stationarity of the DFT+DMFT approach. The self-energy $\Sigma(\omega)$ contains all
Feynman diagrams local to the Fe ion. No downfolding or other approximations were used,
and the calculations are all-electron as implemented in Ref. 67, which is based on Wien2k.\cite{wien2k}
We used the GGA exchange-correlation functional,\cite{PBE} and the quantum impurity
model was solved by the continuous time quantum Monte Carlo impurity solver \cite{CTQMC}
using $U=5.0$ eV and $J=0.72$ eV. Brillouin zone integration is done on the 12$\times$12$\times$6
k-point mesh for the AFM unitcell of LaFeAsO containing 4 Fe atoms. All calculations are done
for $T=150$ K. We consider both paramagnetic (PM) and AFM states at this temperature. The AFM
state is considered to represent the actual stable phase observed in experiments, while the PM state
is also calculated to understand the driving force with which the AFM state is
stabilized from a 'bare' state.

\section{Results}
\subsection{Electronic correlation and magnetic strengths as functions of doping}
Schematic phase diagram of LaFeAsO$_{1-x}$H$_x$ in the $x-T$ space is depicted in Fig.~1(a). The first
AFM phase with a stripe-type order is rapidly suppressed and disappears around $x=0.05$ with the emergence
of the first superconducting phase followed by the adjacent second superconducting phase. Further doping 
initiates the second AFM phase, of the same ordering pattern with the first AFM phase, but with a slightly 
different atomic displacement. \cite{Hiraishi} We perform
the DFT+DMFT calculations to check if this suppression and reappearance of the AFM phase can be 
reproduced. 
To take the electron doping effect into account, we adopt the virtual crystal approximation.\cite{Iimura}
%using a virtual atom with the number of electrons corresponding to the concentration $x$ to replace the
%normal O species.\cite{Iimura}
In addition, the structural change due to the H doping is incorporated by interpolating
both the lattice constants and internal atomic coordinates of the available experimental values
at $x=0$ \cite{Nomura} and $x=0.51$ \cite{Hiraishi} in the PM states with the tetragonal
lattice symmetry. This should be a reasonable approximation
considering the almost linear As height dependence on $x$ observed experimentally,\cite{Iimura}
avoiding the difficulty in the structural optimization of alloy structures within DFT which
would require rather complicated statistical treatment, besides concerns about the general
reliability of DFT in predicting the accurate lattice structure of iron-based superconducting materials.
Using this doping scheme, the static magnetic moment and the local spin susceptibility,
$\chi_{local}(\omega=0)=\int_0^\beta \! \langle S(\tau)S(0) \rangle \mathrm{d}\tau$, are calculated as a function
%which corresponds to the time-averaged local spin-spin correlation, 
of $x$ for the stripe-type AFM phase and shown in Fig.~1(b). The magnetic moment at $x=0$
is estimated to be 0.66 $\mu_B$ with an agreement with the measured value 0.63 $\mu_B$,\cite{Qureshi} and
decreases to the minimum value of 0.12 $\mu_B$ at $x=0.4$, and then exhibits a rapid
increase to reach 0.68 $\mu_B$ at $x=0.6$. The local susceptibility shows a similar behavior with
a minimum at $x=0.3$, implying that the overall magnetic strength is suppressed and then re-enhanced with
the doping. Therefore, we verify that the DFT+DMFT method captures the essential
underlying physics of two separate AFM phases in this material and produces a consistent behavior
of the local magnetic strength, while the complete suppression of magnetic phase and emergence
of the superconducting phase in the intermediate $x$ as shown in Fig.~1(a) could not be properly described within the current
calculation scheme. For comparison, we perform the DFT calculation on the relative stability
of the AFM phase and the magnetic moment as functions of $x$.
For $x=0$, the AFM phase is found to be 180 meV/Fe more stable than the nonmagnetic phase with the magnetic
moment of 2.15 $\mu_B$. Upon increasing $x$, the stability of AFM phase and the magnetic moment
show no discernible change, suggesting that the normal DFT calculation cannot properly describe the observed
evolution of the magnetism.

Then we investigate the underlying mechanism of the doping-induced change of the electronic structure
by considering the electron addition and
structural modification separately. Because the nominal number of valence electrons in a Fe atom
for the undoped material is six, which corresponds to an electron doped system from the half-filled orbitals,
further electron doping should result in the monotonic decrease of the correlation strength.\cite{Medici}
On the other hand, the H doping increases the distance between Fe and surrounding As atoms
%(from 2.41 \AA~to 2.44 \AA~for $x=0$ and 0.51, respectively)
as determined experimentally,\cite{Nomura,Hiraishi} which would lead to the localization
of Fe $d$ orbitals.\cite{Fe-As_distance} To confirm this speculation, we estimate the mass enhancement
$1/Z = 1-\frac{\partial\Sigma(\omega)}{\partial\omega}\rvert_{\omega=0}$ for the two effects separately.
First, we calculate $1/Z$ of the $d_{xy}$ in the PM phase
as a function of $x$ considering only the electron addition effect by fixing the lattice structure
to that of $x=0$ as in Fig.~1(c). Indeed, $1/Z$ monotonically increases with the electron addition.
On the contrary, when only the structural effect is included without extra electron, $1/Z$
monotonically decreases with increasing $x$. When
these two competing effects are combined, $1/Z$ increases overall with the doping, which means
that the localization by the structural modification becomes more dominant at the highly doped system.

This competing effects are also reflected on the magnetic strength of the AFM phase as shown in Fig.~1(d).
When only electron
addition effect is considered, the local susceptibility is found to monotonically decrease with increasing
$x$, while it increases monotonically when only the structural modification is taken into account (with
the number of extra electrons fixed to 0.6), in agreement with the behavior of the mass enhancement factors
in Fig.~1(c). Therefore, we can conclude that the initial suppression and the later re-enhancement of the magnetism
with the H doping originates from the two competing effects : the electron addition and increasing Fe-As
distance which suppresses and enhances the local correlation and hence the local magnetism, respectively.
Our analysis naturally draws attention to the important role of the electron correlation and also the structural effect
in understanding the doping induced phase diagram of this material. Suppression of the magnetism and the existence
of the quantum criticality in phosphorus-doped Ba122 systems BaFe$_2$As$_{2-x}$P$_x$ \cite{Jiang,Rotter,Analytis}
are another set of examples which demonstrate the dramatic effect of the pure
structure modification, where decreased Fe-anion distance was pointed out to cause the suppression of the
magnetism.\cite{Rotter}

\subsection{Fermi surfaces}
We also investigate the evolution of the FS with the doping which is generally considered to be
relevant to the existence of the AFM phase.\cite{Fujiwara,Suzuki} The FSs for three different
doping levels, $x=$0, 0.3, and 0.5, are calculated with both the DFT+DMFT and DFT methods and
displayed in Fig.~2. Starting with a relatively good nesting
between the hole and electrons surfaces at $x=0$ for the DFT case, the doping degrades the nesting
with shrinking the hole surfaces at the $\Gamma$ point and enlarging the electron surfaces at the $M$
point (see Fig.~2(g)-(i)), as the electron doping raises the Fermi level. Our result shows a good agreement with the previous
DFT calculation using the experimentally determined lattice structure,\cite{Iimura} confirming that
our assumption of the linear dependence of the lattice constants and internal atomic coordinates is
reasonable. The DFT+DMFT results shown in Figs.~2(a)-(f) are qualitatively similar, but the $d_{xy}$
hole FS expands compared with the DFT results, because of more correlated nature of the $d_{xy}$
orbital than $d_{xz/yz}$ orbitals as pointed out in the DFT+DMFT study of LiFeAs.\cite{geunsik} The decrease of the
overall spectral weight with the doping and relatively larger incoherence of the $d_{xy}$ surface reflect the larger correlation
at high doping levels and for the $d_{xy}$ orbital (Figs.~2(a)-(c)). Nevertheless, both levels
of the theory indicate the monotonic degradation of the FS nesting with the doping, as already
indicated by previous calculations,\cite{Iimura,Suzuki} manifesting that the FS nesting alone
cannot explain the appearance of the second AFM phase. Again, we conclude that
the electron correlation and many-body effects should be incorporated to understand the doping-induced
evolution of the magnetism.

\subsection{Spin resolved spectral function}
To understand the doping-induced suppression and the reemergence of the magnetism in detail,
we investigate the spin-resolved spectral function of the $d_{xy}$ orbital in the AFM phase as a function of $x$
as shown in Fig.~3. At $x=0$, the majority and minority spin states exhibit a large exchange
splitting reflecting the overall magnetic moment of 0.66 $\mu_B$, with a distinct pseudo-gap feature (a dip
in spectral function) at the Fermi energy induced by the coupling between the electron and hole bands at the
Fermi energy.\cite{Moon1}
With increasing doping level up to $x=0.3$ (see Fig.~3(a)), spectral weights moves from
the peak just above the Fermi level to one below the Fermi energy in the minority
spin channel, suggesting the doped electrons fills the minority spin states. The electron
filling in the minority spin states with the doping naturally leads to the gradual reduction
of the exchange splitting and the magnetic moment, along with the size of the pseudo-gap. On
the other hand, further doping over $x=0.3$ enhances the exchange splitting as shown in
Fig.~3(b), which seems to almost retrace the evolution of the spectral function from $x=0$
to $x=0.3$ in Fig.~3(a). However, there are several noticeable differences as well.
First, the pseudo-gap position is constantly shifting deeper in the valence states
with its size and the magnetic moment increasing with increasing $x$, which indicates
the rise of the Fermi energy as a result of the electron doping. More importantly, doping
over $x=0.3$ develops a shoulder growing with $x$ near -1 eV in the majority spin channel
as indicated by a arrow in Fig.~3(b),
contributing to build up the magnetic moment against the electron filling on the minority
spin states with the doping. The spectral weight piled up in this position results
from many-body effects and hence is incoherent, rather than from the shift of
coherent quasi-particle states. The inset of Fig.~3(b) depicts the imaginary part of
the $d_{xy}$ component of the electron self energy (Im$\Sigma$) along with the spectral function
for $x=0.6$. One can identify a strong peak of the Im$\Sigma$ around -0.7 eV, close
to the $J$ value 0.72 eV adopted in this study, and the shoulder structure of
the spectral function at a nearby position, suggesting that the shoulder structure
originates from the incoherent excitations related to the self energy.
Similar energy scales between this incoherent excitation and $J$ is also consistent with the
suggestion that iron-based superconductors are Hund's metals where $J$ plays more
important role than $U$.\cite{Haule2}

\subsection{Orbital polarization}
As mentioned earlier, the orbital order is of great interest for these materials
regarding the electronic nematicity and also superconductivity itself. Here we compare the
orbital polarization, i.e., the imbalance between $d_{xz/yz}$ orbitals, in the AFM state
for low and high doping cases. For $x=0$ as displayed in Fig.~4(a), $d_{xz}$ and $d_{yz}$
spectral functions show noticeable difference, where the spin polarization is larger for
$d_{yz}$ as well as $d_{xy}$ orbitals while $d_{xz}$ spin splitting is smaller.
On the other hand, for the high doping case of $x=0.5$ in Fig.~4(b),
$d_{xz/yz}$ components of the spectral function becomes much more similar with
each other and now the $d_{xy}$ orbital has the most significant spin polarization.
Indeed, our estimated orbital polarization $(n_{xz}-n_{yz})/(n_{xz}+n_{yz})$ decreases
from 3.9~$\%$ at $x=0$ to 1.5~$\%$ at $x=0.5$ while the magnetic moments for
the two cases are comparable.
Increasing $x$ enhances the crystal field splitting pushing up the $d_{xy}$ level
above $d_{xz/yz}$ level, so that the doped electrons fill $d_{xz/yz}$ orbitals first
rather than the $d_{xy}$ orbital, reducing the imbalance between $d_{xz/yz}$ orbitals
as well as between their spin components.
Meanwhile, besides the less electron filling, the elevated As height
in the high doping case further enhances the electron correlation for the $d_{xy}$ orbital
via the `kinetic frustration' \cite{YinNMAT} compared with the $d_{xz/yz}$ orbitals.
So $d_{xy}$ becomes the most significant component for the local spin fluctuations
in the PM phase and for static magnetic moments in the AFM phase.
Therefore, at high doping levels, strong local magnetism appears mainly from $d_{xy}$
orbital and the orbital polarization from $d_{xz/yz}$ is largely suppressed.

\subsection{Spin excitation spectrum}
So far, we have considered the doping induced evolution of the magnetically ordered state,
and we will conclude our discussion by 
investigating the dynamic spin fluctuations in the PM state, which is more relevant
to the superconductivity, as a function of
the doping. Although $T=150$ K at which the calculation is done is close to the AFM
transition temperature, the spin susceptibility in the PM state is expected to be a smooth 
varying function of $T$ (except at the AFM ordering wave vector for which the susceptibility
diverges at the AFM transition temperature), so we expect qualitatively similar results for other temperatures.
We evaluate the dynamical spin structure factor $S({\bf q},\omega)=
\frac{\chi''({\bf q},\omega)} {1-e^{\hbar\omega/k_BT}}$ using the DFT+DMFT method
as displayed in Fig.~5, where both the one-particle Green's function and the local 
two-particle vertex function are determined {\it ab-initio}.\cite{Hyowon}
For $x=0$, the spin excitation spectrum has strong peaks near the zero energy around
the wave vector ${\bf q}=(\pi,0)$, which corresponds to the magnetic ordering vector
of the AFM phase, and disperses over the path shown in the spectrum, reaching
a maximum energy at the zone boundary ${\bf q}=(\pi,\pi)$, all consistent with
previous results.\cite{Hyowon,Chenglin,Yin2014} As the doping level $x$ increases, the
overall spin excitation spectral weights tend to shift to lower energies as the spin
wave dispersion decreases with the increasing correlation strength. The excitation
near ${\bf q}=(\pi/2,\pi/2)$ noticeably goes down towards the zero energy with the doping,
and a new possible static magnetic order for this wave vector is suggested for $x=0.5$.
However, the intensity of excitations has always the maximum at the conventional AFM
ordering vector ${\bf q}=(\pi,0)$ for all the doping cases, consistent with the
experimentally found second AFM phase for the high doping levels.\cite{Hiraishi}
For $S({\bf q},\omega)$ at ${\bf q}=(\pi+\delta,0)$ which is slightly off the magnetic
ordering vector, as shown in Fig.~5(d), the peak height
near the zero energy is reduced for $x=0.3$ compared with that for $x=0$ indicating the suppressed
tendency towards the static magnetic order, and then it becomes pronounced
again for the higher doping level $x=0.5$ suggesting the re-enhanced magnetism, which
shows a qualitative agreement with the initial decrease and
re-enhancement of the calculated magnetic moment in the AFM phase shown in Fig.~1
and again also with the motivating experiments.\cite{Fujiwara,Hiraishi}
When decomposed by orbitals (see Figs.~5(e)-(g)), the $d_{xz/yz}$ components show a
large anisotropy with the $d_{yz}$ component peak being dominant at $x=0$.
As the doping level increases, the $d_{xz/yz}$ anisotropy keeps decreasing while the $d_{xy}$
component becomes most prominent. The decreasing $d_{xz/yz}$ anisotropy and the enhancement of
the $d_{xy}$ component with the increasing doping is consistent with the features
observed in our result for the AFM phase.

\section{Discussion}
Our results remind us of the indispensable role of the electron correlation
in the iron-based superconducting materials, as well as the impact of the structural
change. The doping-induced evolution of the electronic and magnetic properties
cannot be understood by simply adopting the rigid shift of the Fermi level
or even the self-consistent addition of carriers without taking the structural
effect into account. In addition, a natural view on the doping-induced evolution
of spin and orbital orders can be obtained.
The `ferro-orbital order', which is coupled to the AFM
spin order in the undoped materials, is the lowest-energy configuration for the
nominally half-filled orbitals to maximize the kinetic energy
gain.\cite{Ku}
Both the orbital and spin orders of $d_{xz}$ or $d_{yz}$ are suppressed when the doping
supplies more electrons to these orbitals away from the half-filling.
The $d_{xy}$ orbital, for which the spin order can form but the orbital order is no
longer relevant, becomes the dominant channel for the electron hopping to reduce
the kinetic energy as discussed above. As a result, spin order/fluctuation is
present near the both superconducting domes found in LaFeAsO$_{1-x}$H$_x$ while
orbital order/fluctuation is expected to be strong only near the first superconducting
phase in the lower doping level.
Our results consequently suggests that the spin fluctuation is more closely
related to the superconductivity, at least for
the second superconducting phase in this alloy, while the orbital fluctuation, which
is significant only at low doping levels, might not be a prerequisite for the superconductivity in general.
The enhanced role of the $d_{xy}$ orbital in magnetism
%(larger contribution to the
%local magnetic moment in the AFM phase and also the larger peak in the spin excitation
%spectrum in the PM phase)
is expected to naturally lead to its dominant
role also in the superconductivity with a larger FS hole pocket of this
orbital as shown in Fig.~2. Although the enhanced electron correlation and consequent
stronger spin fluctuation in the higher doping level might contribute to the strong
superconductivity, too strong correlation would be harmful to the superconductivity,
out of several reasons,\cite{YinNMAT} due to the lowered magnon energy scale (Fig.~5)
which is directly coupled to the size of the superconducting gap.
%AFM2-AFM3 crossover...suddenly terminates the superconducting phase..
Further theoretical study which directly attacks the superconductivity as a function
of the doping level will be desirable.

\section{Conclusion}
In summary, by adopting the DFT+DMFT method,
where the local dynamic correlation effect is taken exactly,
we successfully reproduce the hydrogen-doping-induced suppression and
revitalization of the magnetism in LaFeAsO$_{1-x}$H$_x$ which has been recently
established experimentally. Taking the structural modification by the doping into
account along with the carrier addition is found to be essential, as the two factors
induce independent and opposite effects on the electron correlation strength and the
magnetism in this alloy. Doping reduces the orbital imbalance between $d_{xz/yz}$ orbitals 
as well as their magnetic activity,
while the $d_{xy}$ orbital becomes the dominant electron hopping channel with increased
electron correlation and the magnetic strength for high doping levels.
Indispensable role of the electron correlation and detailed atomic structure is identified
in understanding the electronic and magnetic properties, and the magnetism possibly as more fundamental
ingredient in realizing the superconductivity is suggested over the orbital degrees of freedom.

%When decomposed by orbitals, we can identify the features consistent with those
%for the AFM case discussed earlier: with increasing doping, the difference between
%$d_{xz}$ and $d_{yz}$ components decreases while the $d{xy}$ component grows,
%especially for ${\bf q}=(1,0,1)$. These features are more obvious when we plot
%$S({\bf q},\omega)$ at a specific wave vector. For total $S({\bf q},\omega)$ at
%the AFM ordering vector ${\bf q}=(1,0,1)$, the peak height near the zero energy
%is reduced for $x=0.3$ compared with that for $x=0$ indicating the suppressed
%tendency towards the static magnetic order, and then it becomes much more pronounced
%again for the
%higher doping level $x=0.6$ suggesting the reenhanced magnetism, which is consistent
%with the initial decrease and
%reenhancement of our calculated magnetic moment in the AFM phase shown in Fig.~1
%and again also with the motivating experiments\cite{Fujiwara,Hiraishi}.
%When orbitally decomposed, $d_{xz/yz}$ components show a large anisotropy with
%$d_{yz}$ component peak is dominant for the undoped case. As the doping level
%increases, $d_{xz/yz}$ anisotropy keeps decreasing while the $d_{xy}$ component becomes
%most prominent.

\begin{acknowledgments}
This research was supported by Basic Science Research Program through the National Research 
Foundation of Korea (NRF) funded by the Ministry of  Science, ICT and Future Planning (2016R1C1B1014715
and 2015R1D1A1A01059621).
\end{acknowledgments}

%\bibliography{LOFA}

\begin{thebibliography}{72}
\bibitem[{Fe-()}]{Fe-As_distance}
\bibinfo{note}{H doping reduces the lattice constants which seems to be natural
considering the smaller size of H than O species which is replaced. Meanwhile,
the Fe-As distance turns out to increase with the doping, which might be understood
by imagining that the reduced thickness of LaO layers by H doping decompresses
adjacent FeAs layers allowing the Fe-As distance to increase.}
\end{thebibliography}

\newpage

\begin{figure}[tp]
\includegraphics[width=0.85\linewidth]{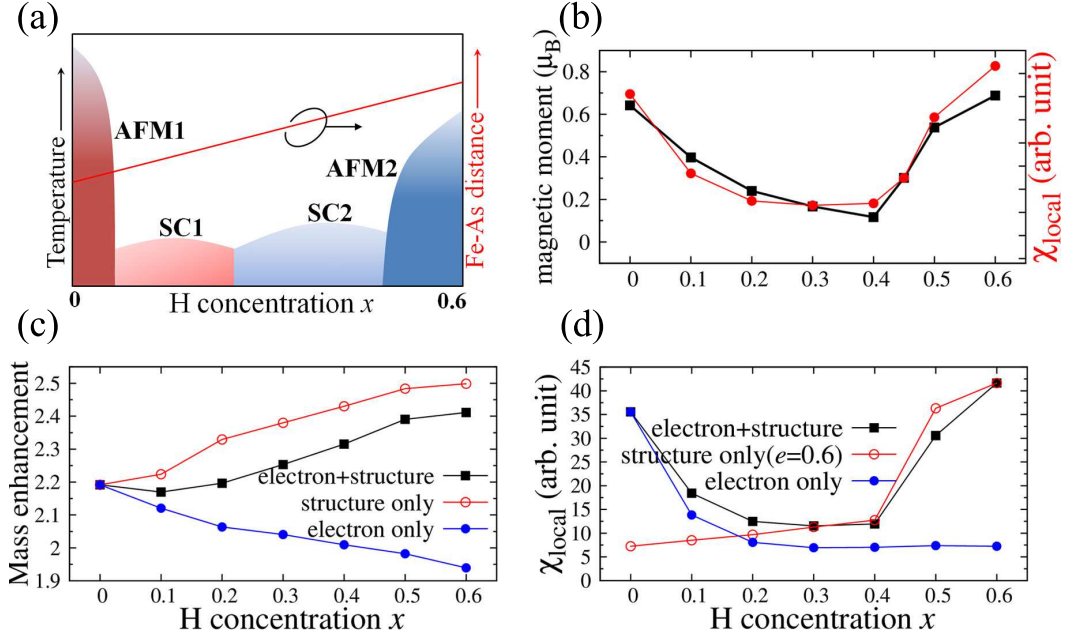}
\caption{(color online) (a) Schematic phase diagram of LaFeAsO$_{1-x}$H$_x$ for the
the hydrogen concentration $x$ and temperature. AFM1 and AFM2 denote
the first and second AFM phases, while SC1 and SC2 represent the
first and second superconducting phases, respectively. Note that the bond length
between Fe and As atoms increases with $x$. (b) Local magnetic
moment and magnetic susceptibility $\chi_{local}(\omega=0)$ in the AFM phase
as a function of $x$, which exhibit minimums around $x=0.3$. (c) Mass enhancement
$1/Z$ is calculated as a function
of $x$ with three different treatments of the doping in the PM phase: both
electron addition and structural change are included (filled squares), only structural
change is allowed while no extra
electron is added (empty circles), and extra electron is added with $x$ while
fixing the structure to that of $x=0$ case (filled circles). (d) Local magnetic
susceptibility as a function of $x$ in the AFM phase for three different treatments
for doping analogously to (c). For the structure only case, we use the number
of extra electrons fixed to that for the $x=0.6$ case.}
\label{fig1}
\end{figure}

%\newpage

\begin{figure}[tp]
\includegraphics[width=0.85\linewidth]{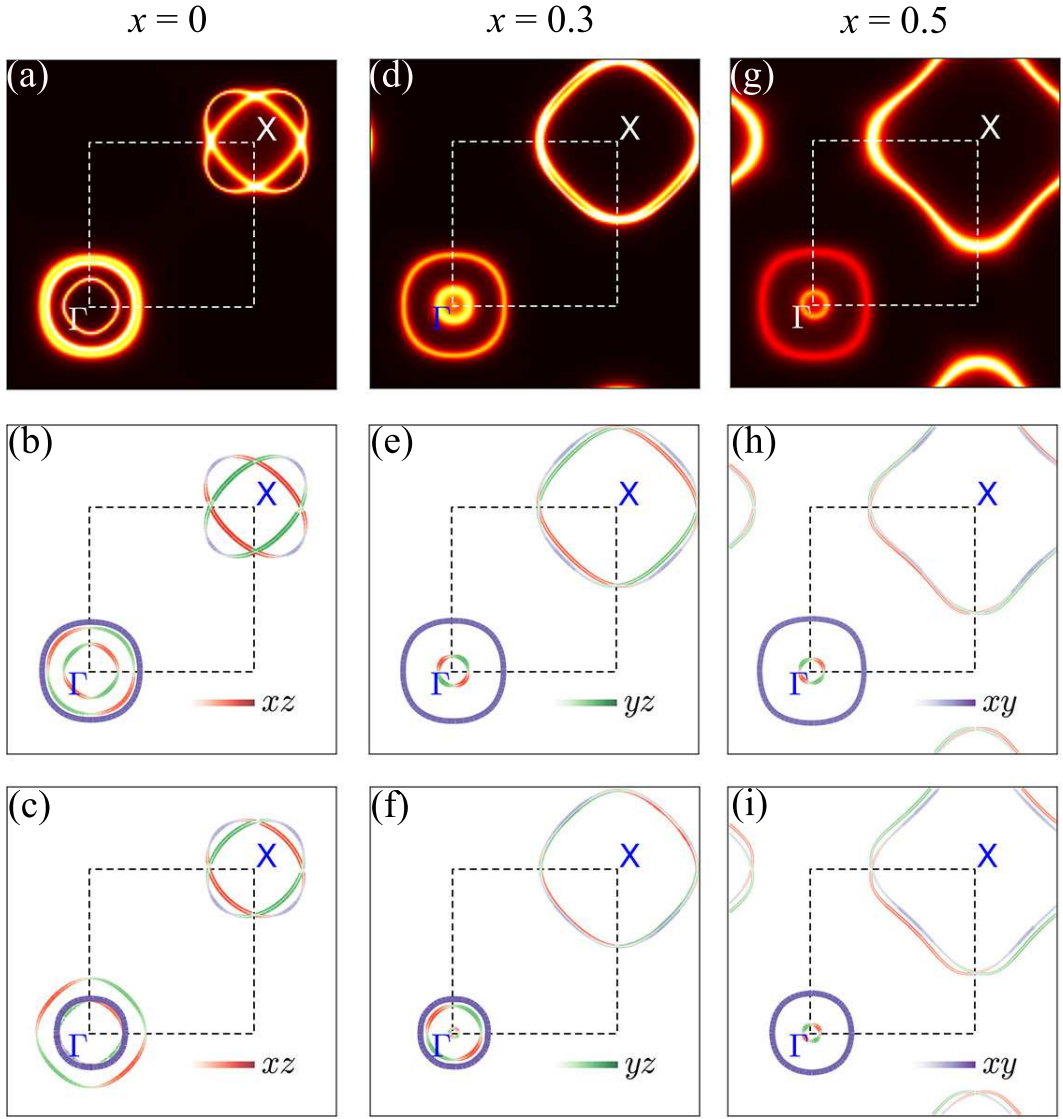}
\caption{(color online) FS for (a)-(c) $x=0$,
(d)-(f) $=0.3$, and (g)-(i) $x=0.5$ calculated on the $k_z=0$ plane of the PM phase.
Top rows are plots of A($k$,$\omega=0$)
from the DFT+DMFT calculations, and middle rows denote their Fe $d$ orbital characters,
while bottom rows are from the usual DFT calculations. Each orbital character
is represented by the depth of the assigned color as well as the thickness of the Fermi
surface line.}
\label{fig2}
\end{figure}

%\newpage
\begin{figure}[tp]
\includegraphics[width=0.85\linewidth]{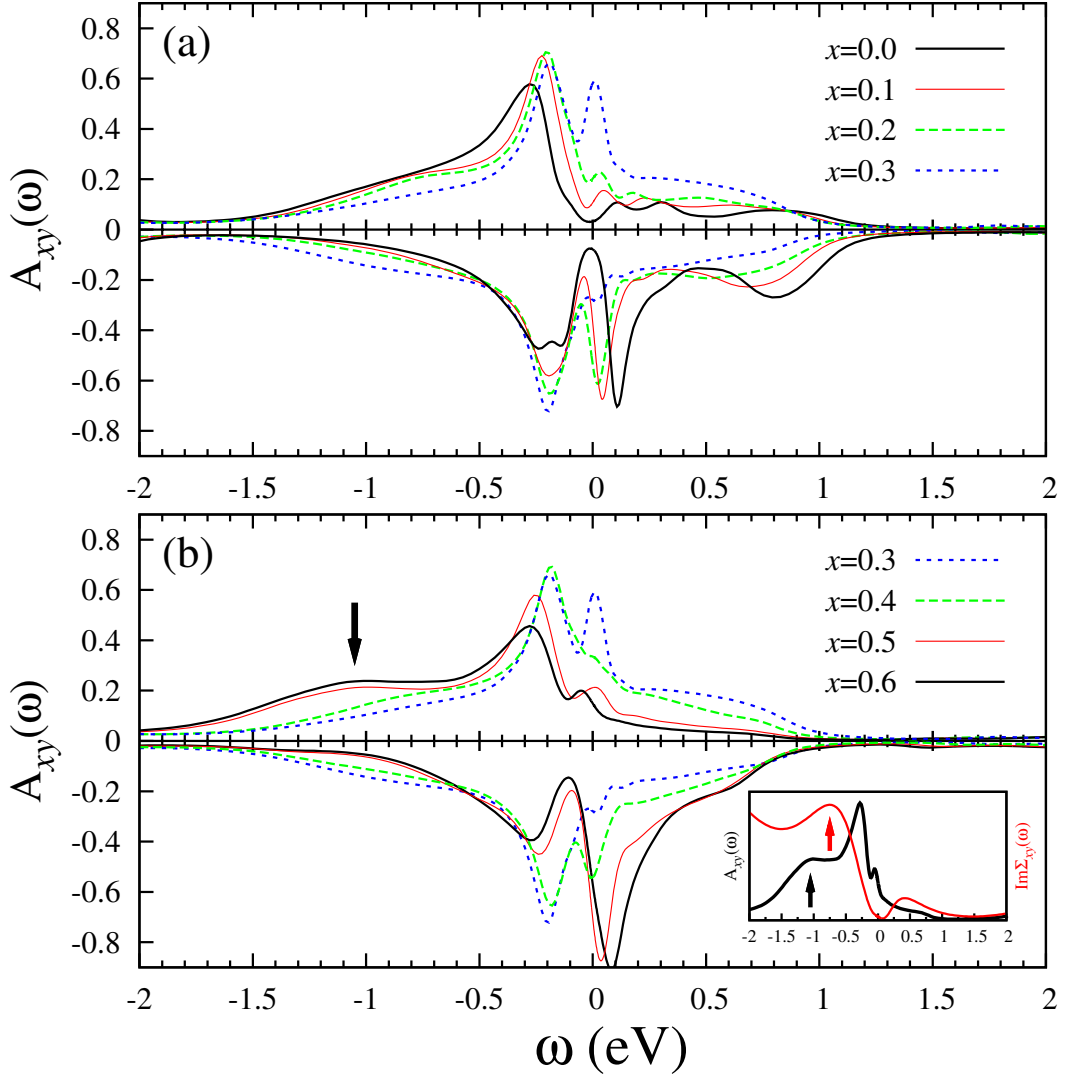}
\caption{(color online)  The spectral function projected onto the Fe $d_{xy}$ orbital of an Fe atom, 
A$_{xy}(\omega)$, with majority
and minority spin channels displayed in positive and negative axes, respectively, in the
AFM phase. (a) From $x=0$ to $x=0.3$, the magnetic moment decreases with decreasing exchange
splitting. (b) From $x=0.4$ to $x=0.6$, the magnetic moment increases. Inset depicts
A$_{xy}(\omega)$ in the black line for majority spin channel of $x=0.6$. The shoulder structure
around -1 eV indicated by a black arrow is correlated with the peak structure in Im$\Sigma_{xy}(\omega)$
in the red line at a slightly higher energy location as indicated by a red arrow.}
\label{fig3}
\end{figure}

\begin{figure}[tp]
\includegraphics[width=0.85\linewidth]{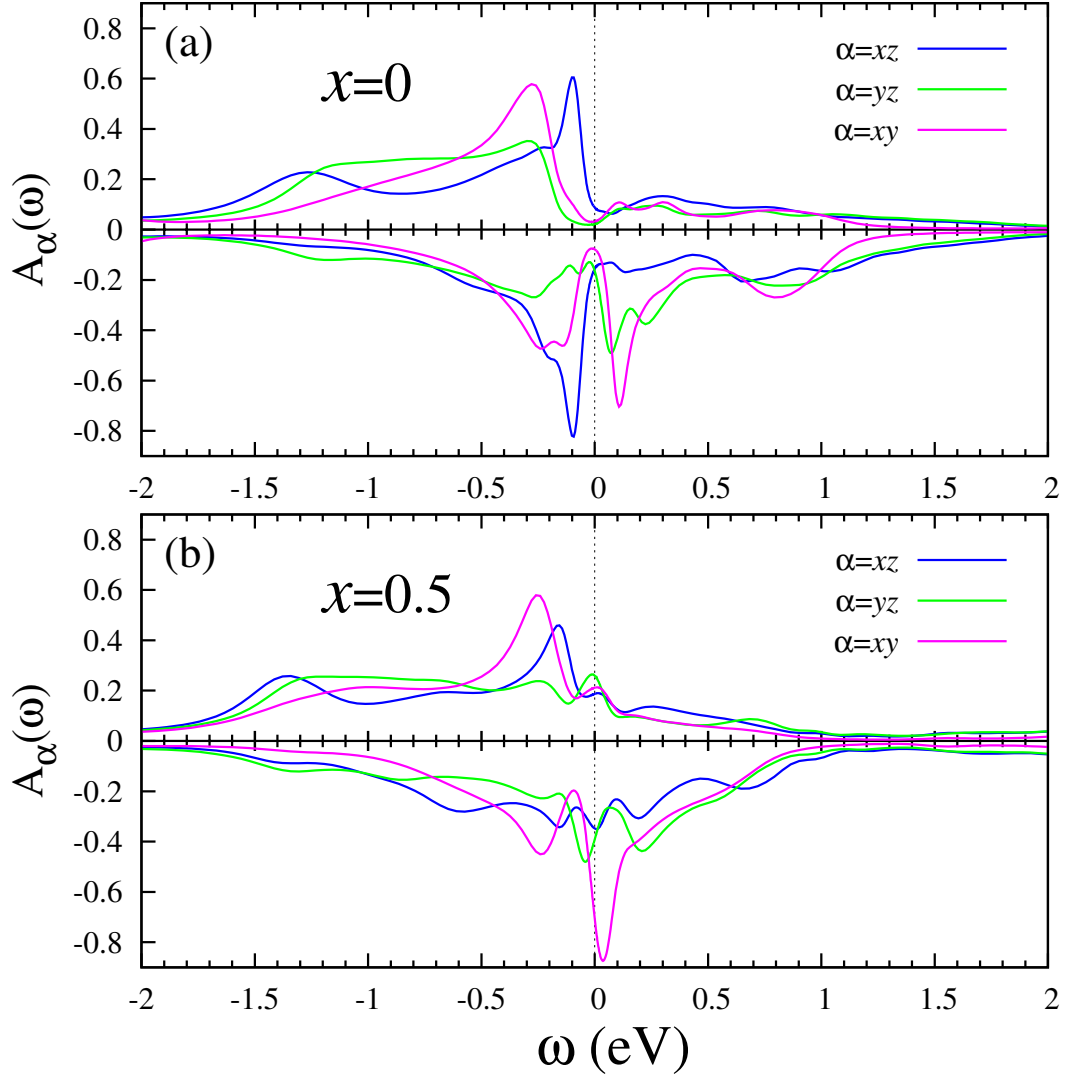}
\caption{(color online)  The spectral function in the two spin channels in the AFM phase, 
projected onto $d_{xz}$, $d_{yz}$, and $d_{xy}$ orbitals in a Fe atom,
for (a) $x=0$ and (b) $x=0.5$. The orbital order which corresponds to
the difference between $xz$ and $yz$ components is suppressed at $x=0.5$ compared with $x=0$.}
\label{fig4}
\end{figure}

%\newpage

\begin{figure}[tp]
\includegraphics[width=0.85\linewidth]{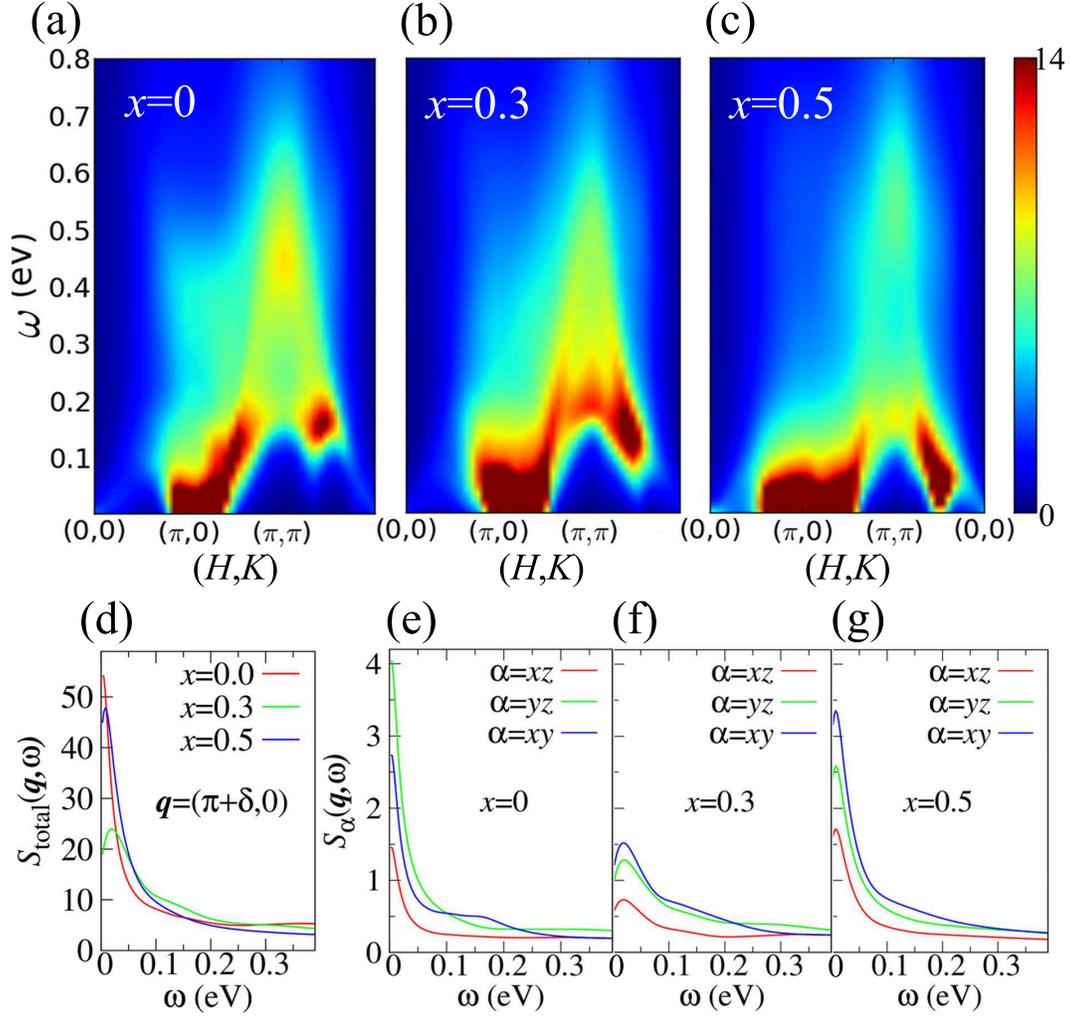}
\caption{(color online) Dynamic spin structure factor $S({\bf q},\omega)$ in the PM state for 
(a) $x=0$, (b) $x=0.3$,
and (c) $x=0.5$ is plotted along the high-symmetry path ($H,K,L=\pi$) in the first Brillouin zone of
the single-iron unit cell, where ($H,K)=(\pi,0$) is the AFM ordering vector. (d) Total $S({\bf q},\omega)$
is plotted as a function of $\omega$ for different $x$ at the fixed momentum $q=(\pi+\delta,0)$, which
is slightly off the AFM ordering vector at which the spin susceptibility diverges below the AFM transition 
temperature. Here we take $\delta= -0.0625\pi$. Orbital decomposed
$S({\bf q},\omega)$ is displayed also at $q=(\pi+\delta,0)$ for (e) $x=0$, (f) $x=0.3$, and (g)
$x=0.5$.}
\label{fig5}
\end{figure}

%\newpage

\end{document}